\documentclass[11pt, a4paper]{article}
\pdfoutput=1
\usepackage{jcappub}

\usepackage{booktabs}
\usepackage{multirow}
\usepackage{amssymb}
\usepackage[export]{adjustbox}
\usepackage{floatrow}
\newfloatcommand{capbtabbox}{table}[][3.5in]
\newfloatcommand{capbfigbox}{figure}[][2.3in]
\usepackage{blindtext}
\usepackage{amsmath}
\usepackage{booktabs}
\usepackage{aas_macros}

\def\lsim{\mathrel{\raise.3ex\hbox{$<$\kern-.75em\lower1ex\hbox{$\sim$}}}}
\def\gsim{\mathrel{\raise.3ex\hbox{$>$\kern-.75em\lower1ex\hbox{$\sim$}}}}

\begin{document}

\hspace*{110mm}{\large \tt FERMILAB-PUB-17-061-A}

\vskip 0.2in

\title{HAWC Observations Strongly Favor Pulsar Interpretations of the Cosmic-Ray Positron Excess}





\author{Dan Hooper,$^{a,b,c}$}\note{ORCID: http://orcid.org/0000-0001-8837-4127}
\emailAdd{dhooper@fnal.gov}
\author{Ilias Cholis,$^{d}$}
\emailAdd{icholis1@jhu.edu}\note{ORCID: http://orcid.org/0000-0002-3805-6478}
\author{Tim Linden$^e$}\note{ORCID: http://orcid.org/0000-0001-9888-0971}
\emailAdd{linden.70@osu.edu}
\author{and Ke Fang$^{f,g}$}\note{ORCID: http://orcid.org/0000-0002-5387-8138}
\emailAdd{kefang@umd.edu}

\affiliation[a]{Fermi National Accelerator Laboratory, Center for Particle Astrophysics, Batavia, IL 60510}
\affiliation[b]{University of Chicago, Department of Astronomy and Astrophysics, Chicago, IL 60637}
\affiliation[c]{University of Chicago, Kavli Institute for Cosmological Physics, Chicago, IL 60637}
\affiliation[d]{Department of Physics and Astronomy, The Johns Hopkins University, Baltimore, Maryland, 21218}
\affiliation[e]{Ohio State University, Center for Cosmology and AstroParticle Physics (CCAPP), Columbus, OH  43210}
\affiliation[f]{University of Maryland, Department of Astronomy, College Park, MD, 20742}
\affiliation[g]{Joint Space-Science Institute, College Park, MD, 20742}

\abstract{Recent measurements of the Geminga and B0656+14 pulsars by the gamma-ray telescope HAWC (along with earlier measurements by Milagro) indicate that these objects generate significant fluxes of very high-energy electrons. In this paper, we use the very high-energy gamma-ray intensity and spectrum of these pulsars to calculate and constrain their expected contributions to the local cosmic-ray positron spectrum.  Among models that are capable of reproducing the observed characteristics of the gamma-ray emission, we find that pulsars invariably produce a flux of high-energy positrons that is similar in spectrum and magnitude to the positron fraction measured by PAMELA and AMS-02. In light of this result, we conclude that it is very likely that pulsars provide the dominant contribution to the long perplexing cosmic-ray positron excess.}

\maketitle

\section{Introduction}
\label{sec:introduction}

Measurements of the cosmic-ray positron fraction by the PAMELA~\cite{Adriani:2010rc} and AMS-02~\cite{Aguilar:2013qda} experiments (as well as HEAT~\cite{Barwick:1997ig}, AMS-01~\cite{Aguilar:2007yf} and Fermi~\cite{FermiLAT:2011ab}) have identified an excess of high-energy positrons relative to the standard predictions for secondary production in the interstellar medium (ISM). This result indicates that significant quantities of $\sim$0.01-1 TeV positrons are produced within the local volume (the surrounding $\sim$\,kpc) of the Milky Way. The origin of these particles has been debated, and possible sources include nearby and young pulsars~\cite{Hooper:2008kg, Yuksel:2008rf, Profumo:2008ms, Malyshev:2009tw, Grasso:2009ma, Linden:2013mqa,Cholis:2013psa}, as well as annihilating dark matter particles~\cite{Bergstrom:2008gr, Cirelli:2008jk, Cholis:2008hb,  Cirelli:2008pk, Nelson:2008hj, ArkaniHamed:2008qn, Cholis:2008qq, Cholis:2008wq, Harnik:2008uu, Fox:2008kb, Pospelov:2008jd, MarchRussell:2008tu, Chang:2011xn,Cholis:2013psa}. This excess could also plausibly arise in nearby supernova remnants, which may be able to produce and accelerate secondary positrons before they escape into the ISM~\cite{Blasi:2009hv, Mertsch:2009ph, Ahlers:2009ae,Cholis:2013lwa,Kachelriess:2011qv, Kachelriess:2012ag,Cholis:2017qlb}. 

From the catalog of known pulsars, Geminga (PSR J0633+1746) and B0656+14 (PSR J0659+1414, thought to be associated with the Monogem supernova remnant) are particularly attractive candidates to account for the observed positron excess. These pulsars are each relatively young (370,000 and 110,000 years, respectively) and are located within a few hundred parsecs of the solar system ($250^{+230}_{-80}$ and $280^{+30}_{-30}$ pc, respectively~\cite{Verbiest:2012kh}). Much of what we know about these and other pulsars is based on gamma-ray observations. Geminga, in particular, is an extremely bright source of GeV-scale emission, thought to originate from the pulsar itself~\cite{2013ApJ...773...77A,2010ApJ...720..272A,2009arXiv0912.5442R,Ackermann:2010aa}. In contrast, observations of Geminga at very high energies reveal a significantly extended morphology. In particular, the Milagro Collaboration has reported the detection of gamma-ray emission at $\sim$\,35 TeV with a flux of $(3.77 \pm 1.07) \times 10^{-16}$ TeV$^{-1}$ cm$^{-2}$ s$^{-1}$ and an extended profile with a full-width at half-maximum of $2.6^{+0.7}_{-0.9}$ degrees~\cite{2009ApJ...700L.127A}.
%
%
Very recently, the HAWC Collaboration has presented their measurements of the Geminga pulsar, confirming its angular extension ($\sim$~$2^{\circ}$ radius) and reporting a flux of $(4.87 \pm 0.68) \times 10^{-14}$ TeV$^{-1}$ cm$^{-2}$ s$^{-1}$ at 7 TeV, with a local spectral index of $-2.23 \pm 0.08$~\cite{Abeysekara:2017hyn} (see also Refs.~\cite{2015arXiv150803497B,2016JPhCS.761a2034C,2015arXiv150907851P}). HAWC has also reported the detection of very high-energy emission from B0656+14, with a similar degree of spatial extension and with a flux and spectral index at 7 TeV of $(2.30 \pm 0.73) \times 10^{-14}$ TeV$^{-1}$ cm$^{-2}$ s$^{-1}$ and $-2.03 \pm 0.14$, respectively~\cite{Abeysekara:2017hyn}.



These observations by HAWC and Milagro allow us to conclude that these pulsars, in fact, deposit a significant fraction of their total spindown power into high-energy leptons. Furthermore, we will show in this paper that the flux of leptons required to explain these observations is roughly equal to that required for the Geminga and B0656+14 pulsars to produce an order one fraction of the positron excess observed by PAMELA and AMS-02. The spectrum, morphology and intensity of the gamma-ray emission measured by HAWC and Milagro leave us with little choice but to conclude that nearby pulsars are likely to be the dominant source of the observed cosmic-ray positrons.

\section{Inverse Compton Scattering of Very High-Energy Electrons and Positrons Near Pulsars}

The gamma-ray emission observed from Geminga and B0656+14 by HAWC and Milagro is almost certainly generated through the inverse Compton scattering of very high-energy leptons. The angular extension of this signal rules out other scenarios, with the possible exception of pion production. A pion production origin, however, would require an unrealistically large quantity ($\gsim$\,$10^{46}$ erg) of $\mathcal{O}(10^2)$ TeV protons to be confined to the region surrounding Geminga for $\gsim$\,$10^5$ years. Such a scenario could also be constrained to some degree by the lack of TeV neutrinos detected by the IceCube experiment~\cite{Aartsen:2014cva}.

To study the diffusion and energy losses of electrons and positrons produced in nearby pulsars, we utilize the standard propagation equation:
\begin{eqnarray}
\frac{\partial{}}{\partial{t}}\frac{dn_e}{dE_e}(E_e,r,t) &=&  \vec{\bigtriangledown} \cdot \bigg[D(E_e) \vec{\bigtriangledown} \frac{dn_e}{dE_e}(E_e,r,t) - \vec{v}_c \,  \frac{dn_e}{dE_e}(E_e,r,t) \bigg] \\
&+& \frac{\partial}{\partial E_e} \bigg[\frac{dE_e}{dt}(r) \, \frac{dn_e}{dE_e}(E_e,r,t)    \bigg]  + \delta(r) Q(E_e,t), \nonumber
\label{diffusionlosseq}
\end{eqnarray}
where $dn_e/dE_e$ is the differential number density of electrons/positrons at a distance $r$ from the pulsar, $D$ is the diffusion coefficient, $\vec{v}_c$ is the convection velocity, and the source term $Q$ describes the spectrum and time profile of electrons/positrons injected into the ISM. Energy losses are dominated by a combination of inverse Compton and synchrotron losses, and are given by~\cite{Blumenthal:1970gc}:
%
%
%
%
\begin{eqnarray}
-\frac{dE_e}{dt}(r) &=& \sum_i \frac{4}{3}\sigma_T \rho_i(r) S_i(E_e) \bigg(\frac{E_e}{m_e}\bigg)^2 + \frac{4}{3}\sigma_T \rho_{\rm mag}(r) \bigg(\frac{E_e}{m_e}\bigg)^2  \\
&\equiv& b(E_e,r) \,  \bigg(\frac{E_e}{{\rm GeV}}\bigg)^2,
\end{eqnarray}
where $\sigma_T$ is the Thomson cross section and
\begin{eqnarray}
b(r) \approx  1.02 \times 10^{-16} \, {\rm GeV}/{\rm s} \, \times \bigg[ \sum_i \frac{\rho_{i}(r)}{{\rm eV}/{\rm cm}^3} \, S_{i}(E_e) + 0.224 \,\bigg(\frac{B(r)}{3\, \mu \rm{G}}\bigg)^2 \bigg].
\end{eqnarray}
The sum in this expression is carried out over the various components of the radiation backgrounds, consisting of the cosmic microwave background (CMB), infrared emission (IR), starlight (star), and ultraviolet emission (UV). Throughout our analysis, we adopt the following parameters: $\rho_{\rm CMB}=0.260$ eV/cm$^3$, $\rho_{\rm IR}=0.60$ eV/cm$^3$, $\rho_{\rm star}=0.60$ eV/cm$^3$, $\rho_{\rm UV}=0.10$ eV/cm$^3$, $\rho_{\rm mag}=0.224$ eV/cm$^3$ (corresponding to $B=3\,\mu$G), and $T_{\rm CMB} =2.7$ K, $T_{\rm IR} =20$ K, $T_{\rm star} =5000$ K and $T_{\rm UV} =$20,000 K. For low to moderate electron energies, these parameters correspond to a value of $b \simeq 1.8 \times 10^{-16}$ GeV/s. At very high energies ($E_e \gsim m^2_e/2T$), however, inverse Compton scattering is further suppressed by the following factor:
\begin{equation}
S_i (E_e) \approx \frac{45 \, m^2_e/64 \pi^2 T^2_i}{(45 \, m^2_e/64 \pi^2 T^2_i)+(E^2_e/m^2_e)}.
\end{equation}
To solve Eq.~\ref{diffusionlosseq}, we calculate the distribution of the electrons and positrons that were emitted a time $t$ ago, and then sum the contributions produced over different periods of time. Considering an injected spectrum of the form $Q(E_e,t) =  \delta(t) Q_0 E^{-\alpha} \exp(-E_e/E_c)$, the solution to Eq.~\ref{diffusionlosseq} (neglecting convection) is given by:
%
%
%
\begin{eqnarray}
\frac{dn_e}{dE_e}(E_e,r,t) =\frac{Q_0 \, E^{2-\alpha}_0}{8\pi^{3/2}E_e^2 \,L^3_{\rm dif}(E_e,t)} \, \exp\bigg[\frac{-E_0}{E_c}\bigg] \, \exp\bigg[\frac{-r^2}{4L^2_{\rm dif}(E_e,t)}\bigg],
\label{solution}
\end{eqnarray}
where $E_0 \equiv E_e/(1-E_e b t)$ is the initial energy of an electron that has an energy of $E_e$ after a time $t$, and the diffusion length scale is given by:
\begin{eqnarray}
L_{\rm dif}(E_e,t) &\equiv&  \bigg[   \int^{E_e}_{E_0} \frac{D(E')}{-dE_e/dt(E')} dE'   \bigg]^{1/2},\\
&=& \bigg[\frac{D_0}{b (E_e/{\rm GeV})^{1-\delta} (1-\delta)} \bigg(1-(1-E_e b t)^{1-\delta}\bigg)\bigg]^{1/2}.
\end{eqnarray}
In the last step we had adopted a parameterization of $D(E_e) = D_0 E_e^{\delta}$ for the diffusion coefficient. Note that for $E_e b t > 1$, there are no electrons/positrons of energy $E_e$ and the contribution to $dn_e/dE_e$ is set to zero. 

To account for the time profile of the electrons and positrons injected from a given pulsar, we adopt a function proportional to the spin-down power (the rate at which the pulsar loses rotational kinetic energy through magnetic dipole braking)~\cite{Gaensler:2006ua}:
\begin{eqnarray}
\label{edot}
\dot{E} &=& -\frac{8\pi^4 B^2 R^6}{3 c^3 P(t)^4} \\
&\approx& 1.0 \times 10^{35} \, {\rm erg}/{\rm s} \times \bigg(\frac{B}{1.6 \times 10^{12} \, {\rm G}}\bigg)^2 \, \bigg(\frac{R}{15 \,{\rm km}}\bigg)^6 \, \bigg(\frac{0.23 \, {\rm s}}{P(t)}\bigg)^4, \nonumber
\end{eqnarray}
where $B$ is the strength of the magnetic field at the surface of the neutron star, $R$ is the radius of the neutron star, and the rotational period evolves as follows:
\begin{eqnarray}
P(t) = P_{_0} \, \bigg(1+\frac{t}{\tau}\bigg)^{1/2},
\end{eqnarray}
where $P_{_0}$ is the initial period, and $\tau$ is the spindown timescale:
\begin{eqnarray}
\label{tau}
\tau&=&\frac{3c^3 I P_{_0}^2}{4\pi^2B^2 R^6} \\
&\approx& 9.1 \times 10^3 \,{\rm years} \, \bigg(\frac{1.6\times 10^{12}\,{\rm G}}{B}\bigg)^2\,\bigg(\frac{M}{1.4 \, M_{\odot}}\bigg) \, \bigg(\frac{15 \, {\rm km}}{R}\bigg)^4 \, \bigg(\frac{P_{_0}}{0.040 \, {\rm sec}}\bigg)^2. \nonumber 
\end{eqnarray}
In Eqs.~\ref{edot} and~\ref{tau}, we had adopted benchmark values for the neutron star's magnetic field, radius and mass, chosen to match Geminga's observed period and its rate of change.

\begin{figure}
\includegraphics[width=3.80in,angle=0]{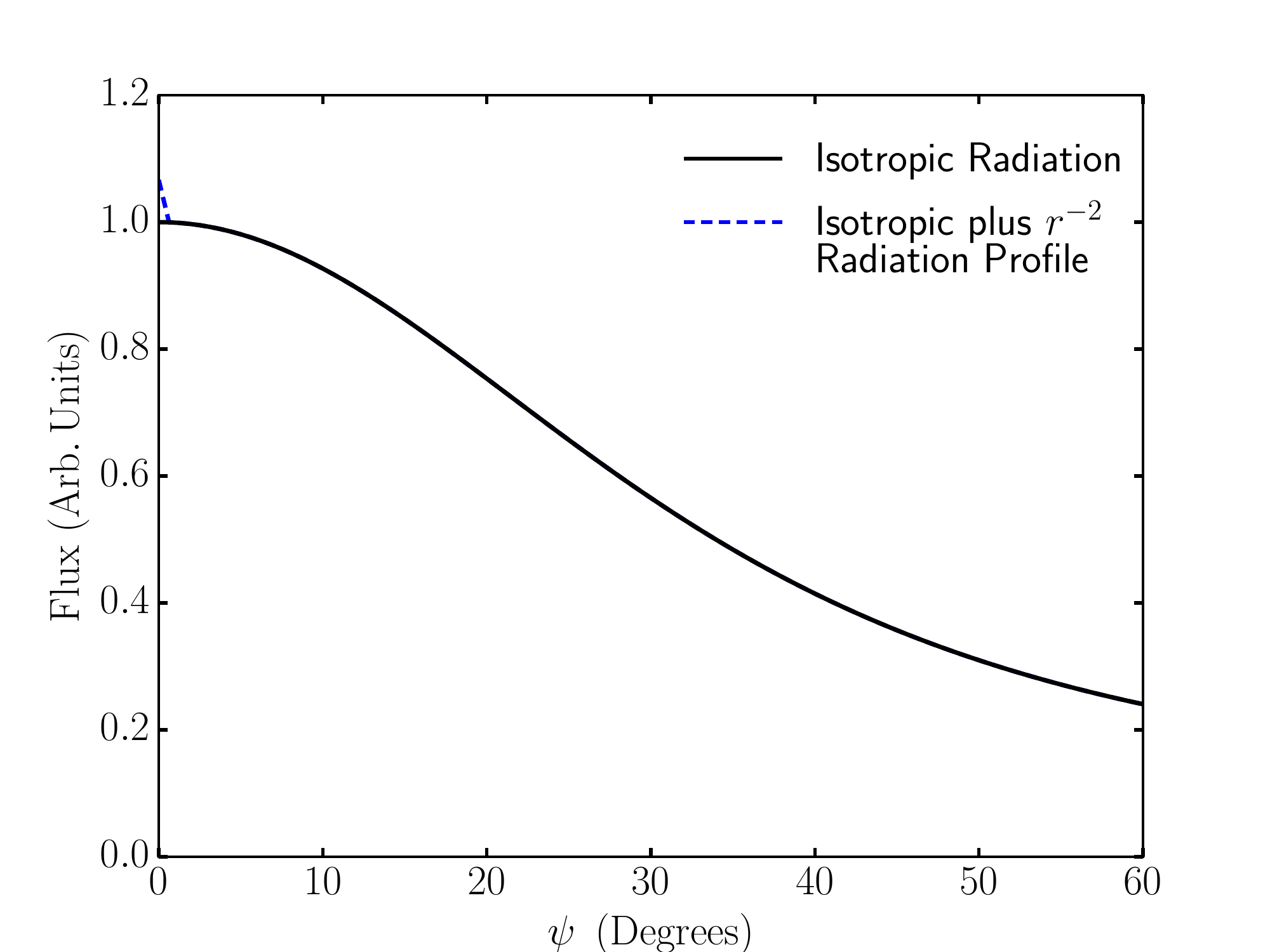}
\caption{The angular distribution of the flux of inverse Compton emission (per solid angle) from 35 TeV electrons (corresponding to photons in the approximate energy range measured by Milagro and HAWC) from the Geminga pulsar. Here, we have adopted diffusion and energy loss parameters which correspond to the conditions found in the ISM, $D_0=2\times 10^{28}$ cm$^2/$s, $\delta=0.4$, and  $b=1.8\times 10^{-16}$ GeV/s, and spectral parameters as given by $\alpha=1.5$, $E_c=100$ TeV. The solid black line represents the angular profile predicted assuming an isotropic radiation distribution, whereas the dashed blue line (visible in the upper left corner) also includes a contribution from a population of radiation which is distributed according to an $r^{-2}$ profile, normalized to the total spin-down power of Geminga. In either case, the predicted profile is dramatically broader than the $\sim$$2^{\circ}$ extension reported by both Milagro and HAWC.}
\label{extrad}
\end{figure}

At energies within the range measured by MILAGRO and HAWC, inverse Compton scattering yields photons with energies not very far below that of the incident electrons and positrons, $E_{\gamma} \sim E_e$. Adopting this approximation, the angular profile of gamma rays generated through Inverse Compton scattering is given by:
\begin{eqnarray}
\Phi_{\gamma}(E_\gamma=E_e,\psi) &\propto& \int \dot{E} \, dt \, \int_{\rm los}  \frac{dn_{e}}{dE_e}(E_e,r,t)  \, \rho_{\rm rad}(r) \, dl \nonumber \\
&\propto&  \int \frac{\tau^2}{(t+\tau)^2} \, dt \int_{\rm los}  e^{-r^2/4L^2_{\rm dif}(E_e,t)}   \, \rho_{\rm rad}(r) \, dl, 
\label{profile}
\end{eqnarray}
where $\psi$ is the angle observed away from the pulsar, and $r^2=l^2+d^2-2ld \cos \psi$, where $d$ is the distance between the pulsar and the observer. If we adopt a uniform distribution of radiation in the vicinity of the pulsar, this reduces to a profile of the form $\Phi_{\gamma}(\psi) \propto \exp[-d^2 \sin^2 \psi/4L_{\rm dif}^2(E_e,t)]$. Observations of Geminga by both Miligro and HAWC indicate that the very high energy gamma-ray emission from this source is extended over a region of a few degrees across the sky. This in turn requires a diffusion length given by $L_{\rm dif}(E_e) \simeq (250 \,{\rm pc}) \, \sin (0.5\times 2.6^{\circ})/2(\ln(2))^{1/2} \simeq 2.6 \, {\rm pc}$. In contrast, adopting parameters appropriate for the ISM ($D_0\simeq 2\times 10^{28}$ cm$^2/$s, $\delta \simeq 0.4$, $b=1.8\times 10^{-16}$ GeV/s), we find
\begin{eqnarray}
L_{\rm dif}(E_e,t) \simeq 200 \, {\rm pc} \,\, \bigg(\frac{35\, {\rm TeV} }{E_e}\bigg)^{0.3} \, \bigg(1-(1-E_e b t)^{0.6}\bigg)^{1/2}.
\end{eqnarray}
Assuming conditions for cosmic ray transport that are similar to those found in the ISM, this calculation shows that we should have expected the Inverse Compton emission observed at very high-energies to be extended over a scale of $\sim$~$60^{\circ}$, dramatically more than the $\sim$~$2^{\circ}$ extension reported by both Milagro and HAWC (see Fig.~\ref{extrad}).

To resolve this puzzle, one might be tempted to consider the possibility that the pulsar is surrounded by a dense radiation field, which intensifies the resulting inverse Compton emission from the surrounding parsecs. The problem with this scenario, however, is that there is not nearly enough power available to generate the required density of radiation. More quantitatively, in order for a $r^{-2}$ profile of radiation to exceed the energy density of the CMB at a distance of 1 parsec from the pulsar would require an amount of energy equivalent to more than ten times the total spin-down power of Geminga. Adopting an extreme benchmark in which 100\% of Geminga's energy budget is transferred into radiation, the profile of Inverse Compton emission is altered only very modestly; by less than 10\% at $\psi = 1^{\circ}$ (see Fig.~\ref{extrad}). Based on these considerations, it does not appear that local concentrations of radiation play a significant role in explaining the angular extent of the gamma-ray emission observed from Geminga or B0656+14.

\begin{figure}
\includegraphics[width=3.80in,angle=0]{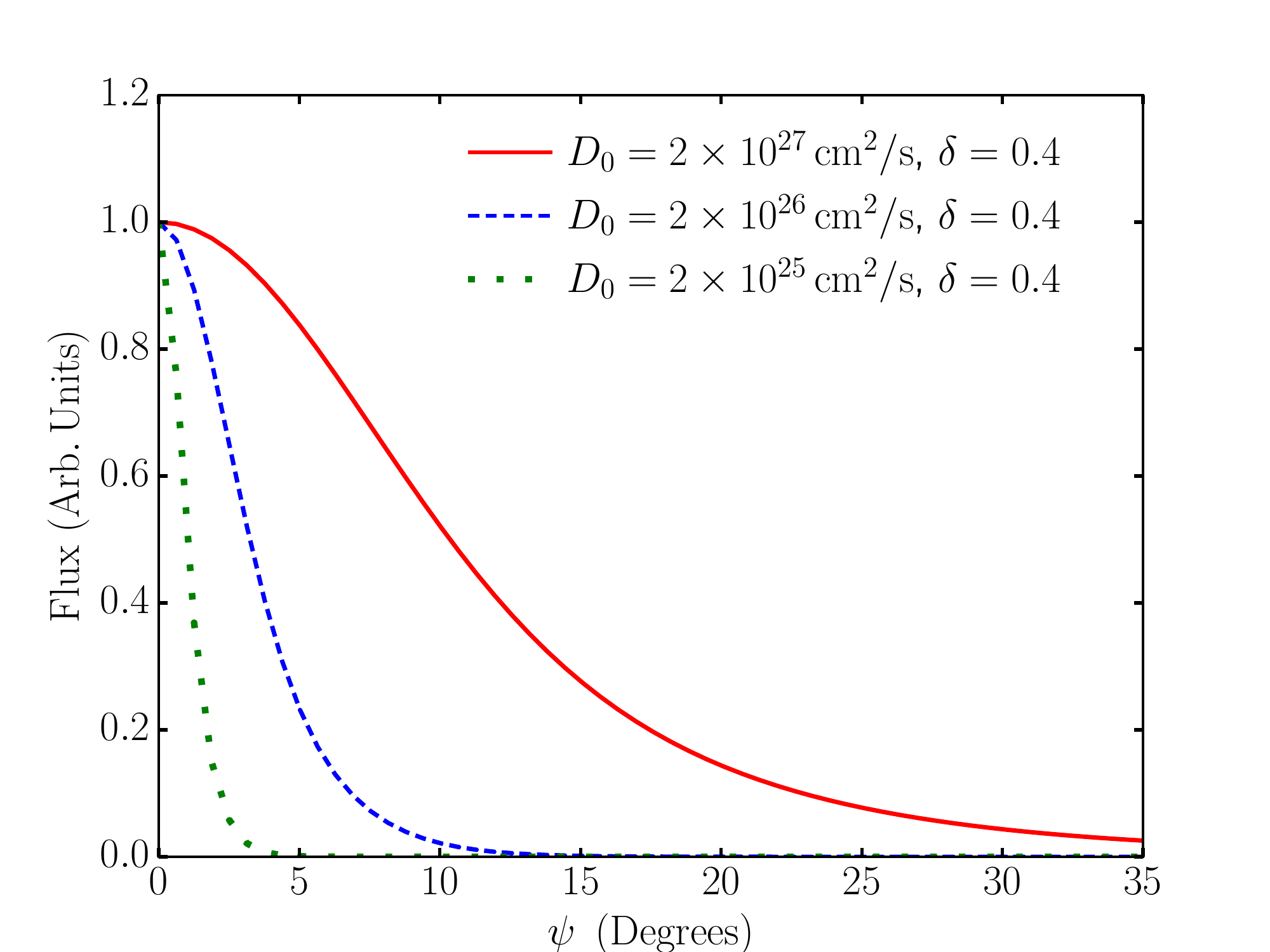}
\caption{The angular distribution of the flux of inverse Compton emission (per solid angle) from 35 TeV electrons from the Geminga pulsar, for different choices of the diffusion parameters, $D_0$, $\delta$ (such that $D=D_0 (E_e/{\rm GeV})^{\delta}$). We have adopted an energy loss parameter of $b=1.8\times 10^{-16}$ GeV/s and spectral parameters of $\alpha=1.5$ and $E_c=100$ TeV. To accommodate the $\sim$$2^{\circ}$ extension reported by Milagro and HAWC, Geminga must be enclosed within an environment with a diffusion coefficient that is $\sim$\,$10^3$ times smaller than that corresponding to the typical conditions of the ISM.}
\label{ext}
\end{figure}

A more likely solution is that the conditions that dictate cosmic-ray diffusion around Geminga and B0656+14 are very different from those found elsewhere in the ISM, leading energetic leptons to escape from the surrounding regions much slower. In order to accommodate the observed extension in this way, we require that these particles only diffuse a distance of a few parsecs before losing most of their energy. For an energy loss time of 5000 years (corresponding to $E_{e}=35$ TeV and $b=1.8\times 10^{-16}$ GeV/s), this requires a diffusion coefficient of $D \sim10^{27}$ cm$^2$/s (see Fig.~\ref{ext}). Although this is significantly smaller than that found in the bulk of the ISM, we note that it is similar to that predicted for standard Bohmian diffusion, $D_{\rm Bohm} = r_L c/3 \approx 1.2 \times 10^{27}$ cm$^2/$s $\times \, (E_e/35 \,{\rm TeV})(\mu {\rm G}/B)$.

%
%
%

If pulsars such as Geminga are typically surrounded by a region with inefficient diffusion ($D \ll D_{\rm ISM}$), the volume of such regions must be fairly small to avoid conflicting with secondary-to-primary ratios in the cosmic-ray spectrum as measured at Earth. In particular, if such regions have a typical radius of $r_{\rm region}$, then such regions will occupy roughly the following fraction of the volume of the Milky Way's disk:
\begin{eqnarray}
f &\sim& \frac{N_{\rm region} \times \frac{4\pi}{3} r^3_{\rm region}}{\pi R^2_{\rm MW} \times 2 z_{\rm MW}} \\
&\sim& 0.25 \times \bigg(\frac{r_{\rm region}}{100 \, {\rm pc}}\bigg)^3 \, \bigg(\frac{\dot{N}_{\rm SN}}{0.03 \, {\rm yr}^{-1}}\bigg) \, \bigg(\frac{\tau_{\rm region}}{10^6 \, {\rm yr}}\bigg) \, \bigg(\frac{20 \, {\rm kpc}}{R_{\rm MW}}\bigg)^2 \, \bigg(\frac{200 \, {\rm pc}}{z_{\rm MW}}\bigg), \nonumber
\end{eqnarray}
where $\dot{N}_{\rm SN}$ is the rate at which new pulsars appear in the Galaxy, $\tau_{\rm region}$ is the length of time that such regions persist, and $N_{\rm region} = \dot{N}_{\rm SN} \times \tau_{\rm region}$ is the number of such regions present at a given time. The quantities $R_{\rm MW}$ and $z_{\rm MW}$ denote the radius and half-width of the Galaxy's cylindrical disk. Combined with Milagro and HAWC observations of Geminga and B0656+14, these considerations suggest $5\, {\rm pc} \lsim r_{\rm region} \lsim 50\,{\rm pc}$, for which there will be little impact on the observed secondary-to-primary ratios (other than the positron fraction).

\begin{figure}
\includegraphics[width=3.08in,angle=0]{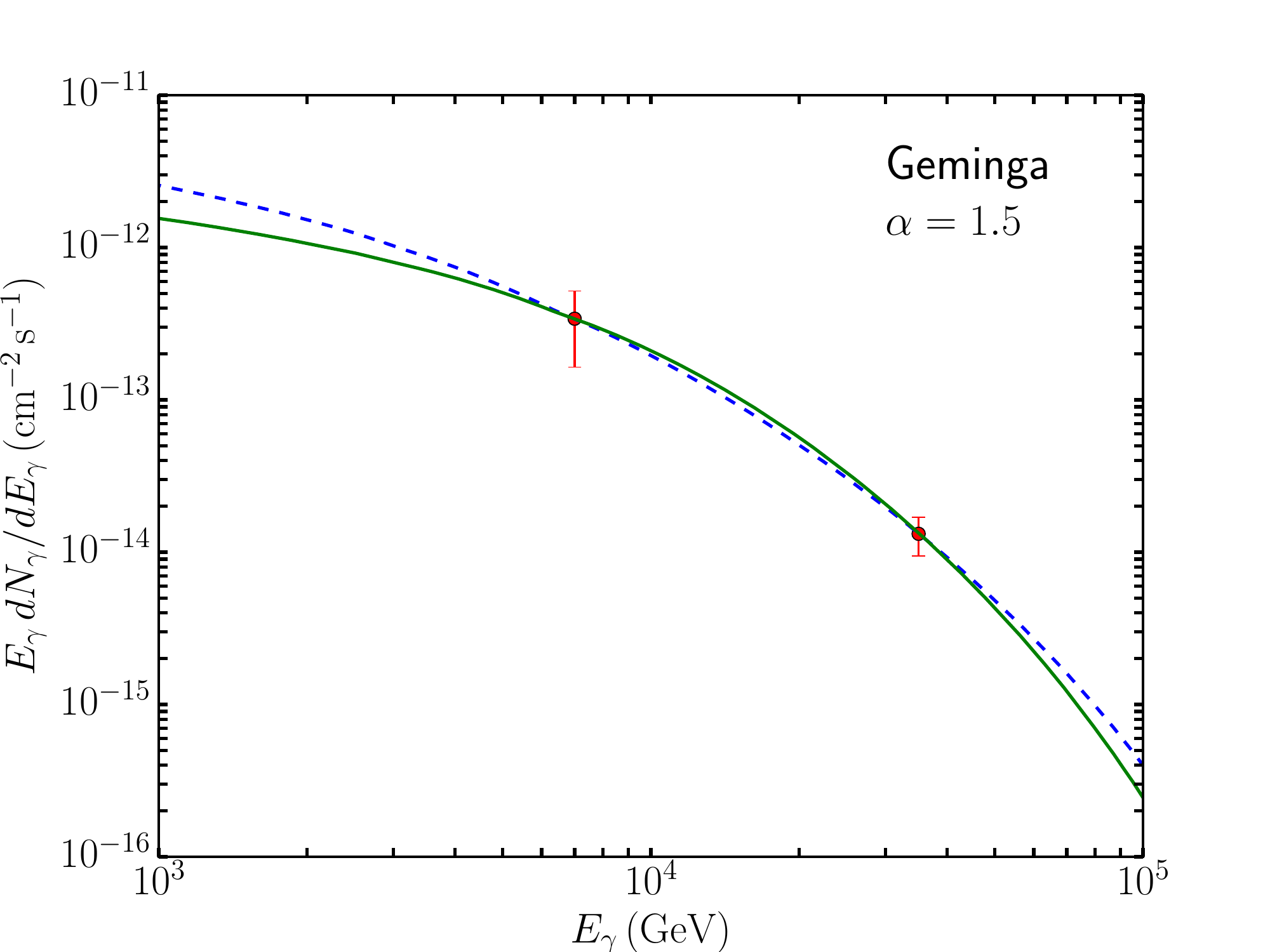}
\hspace{-0.5cm}
\includegraphics[width=3.08in,angle=0]{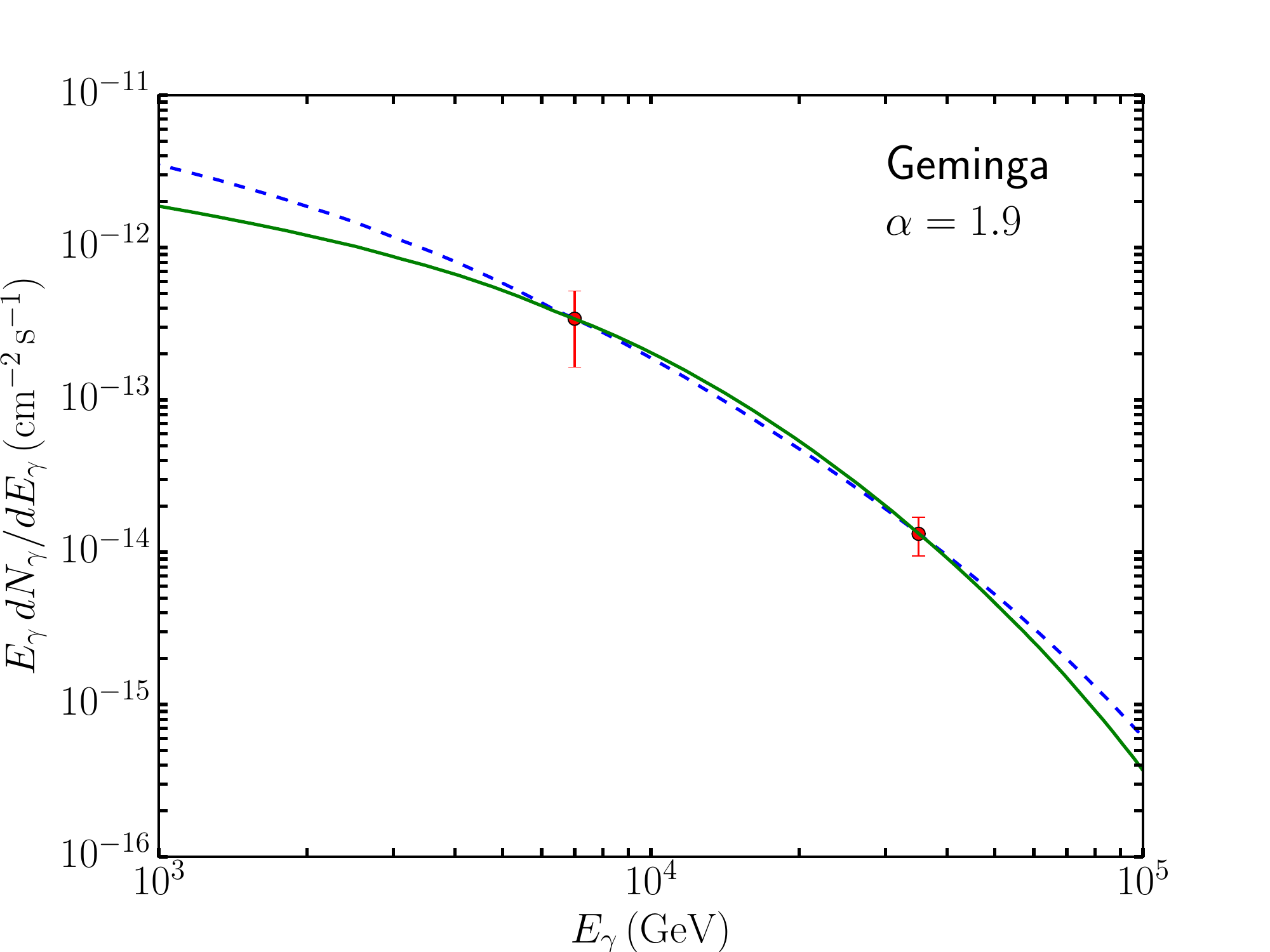}
%
\caption{The gamma-ray spectrum from the region surrounding Geminga, compared to measurements by HAWC and Milagro (shown at 7 and 35 TeV, respectively). In the left (right) frame, we adopt a spectrum of injected electrons with an index of $\alpha=1.5$ (1.9), and have in each case selected a value of $E_c$ that provides the best-fit to the combination of these two measurements. The blue-dashed (solid green) curves correspond to a case with weak convection, $v_c =$ 55.4 km/s $\times \,(r_{\rm region}/10 \, {\rm pc})$ (strong convection, $v_c =$ 554 km/s $\times \,(r_{\rm region}/10 \, {\rm pc})$). As discussed in the text, the case of weak convection is disfavored by the spectral index reported by the HAWC Collaboration.}
\label{gammageminga}
\end{figure}

In Fig.~\ref{gammageminga}, we plot the gamma-ray spectrum from the region surrounding Geminga, compared to the measurements by HAWC and Milagro (at 7 and 35 TeV, respectively). In performing this calculation, we utilized the full differential cross section for inverse Compton scattering~\cite{Blumenthal:1970gc}. In the left (right) frame, we have adopted a spectrum of injected electrons with an index of $\alpha=1.5$ (1.9), and in each case selected a value of $E_c$ that provides the best-fit to the combination of these two measurements. We have also allowed for the possibility that convective winds play a role in cosmic-ray transport~\cite{Yuksel:2008rf}, moving particles away from the pulsar at a velocity given by either $v_c =$ 55.4 km/s $\times \,(r_{\rm region}/10 \, {\rm pc})$ (blue dashed) or $v_c = 554$ km/s $\times \,(r_{\rm region}/10 \, {\rm pc})$ (green solid). In these four cases, the best-fits were found for $E_c=44$ TeV ($\alpha=1.5$, low convection), $E_c=35$ TeV ($\alpha=1.5$, high convection), $E_c=67$ TeV ($\alpha=1.9$, low convection), and $E_c=49$ TeV ($\alpha=1.9$, high convection). In each case, convection dominates over diffusion in transporting cosmic rays out of the region surrounding the pulsar.

In addition to their flux measurement, the HAWC Collaboration has also reported a value of $-2.23 \pm 0.08$ for Geminga's spectral slope at 7 TeV. Among the models shown in Fig~\ref{gammageminga}, those with a low convection velocity ($v_c =$ 55.4 km/s $\times \,(r_{\rm region}/10 \, {\rm pc})$) predict spectral slopes at 7 TeV of $-2.47$ ($\alpha=1.5$) or $-2.59$ ($\alpha=1.9$). Such values are highly inconsistent with that reported by HAWC. In contrast, for those models with a higher degree of convection ($v_c = 554$ km/s $\times \,(r_{\rm region}/10 \, {\rm pc})$), we instead predict a spectral slope of $-2.23$ ($\alpha=1.5$) or $-2.32$ ($\alpha=1.9$), in excellent agreement with HAWC's measurement. This favors scenarios in which convection plays a very important role in the transport of high-energy leptons, especially among those with energies below $\sim$10 TeV. We lastly note that among this range of models, between 7.2\% and 29\% of Geminga's total current spindown power is being deposited into electron-positron pairs with $E_e > 10$ GeV.



\section{Implications for the Cosmic-Ray Positron Excess}

Although the angular extent of the emission observed from Geminga and B0656+14 by HAWC and Milagro indicates that very high-energy ($E_e \gsim 10$ TeV) leptons are effectively trapped within a few parsecs of these sources, the same fate need not be experienced by lower energy electrons and positrons. In particular, even a modest degree of convection (i.e. the streaming of particles away from the source at a constant velocity) can remove sub-TeV leptons from the region before they lose a substantial fraction of their energy, while higher energy leptons lose the vast majority of their energy to inverse Compton scattering before escaping the same region~\cite{Yuksel:2008rf}.

\begin{figure}
\includegraphics[width=3.08in,angle=0]{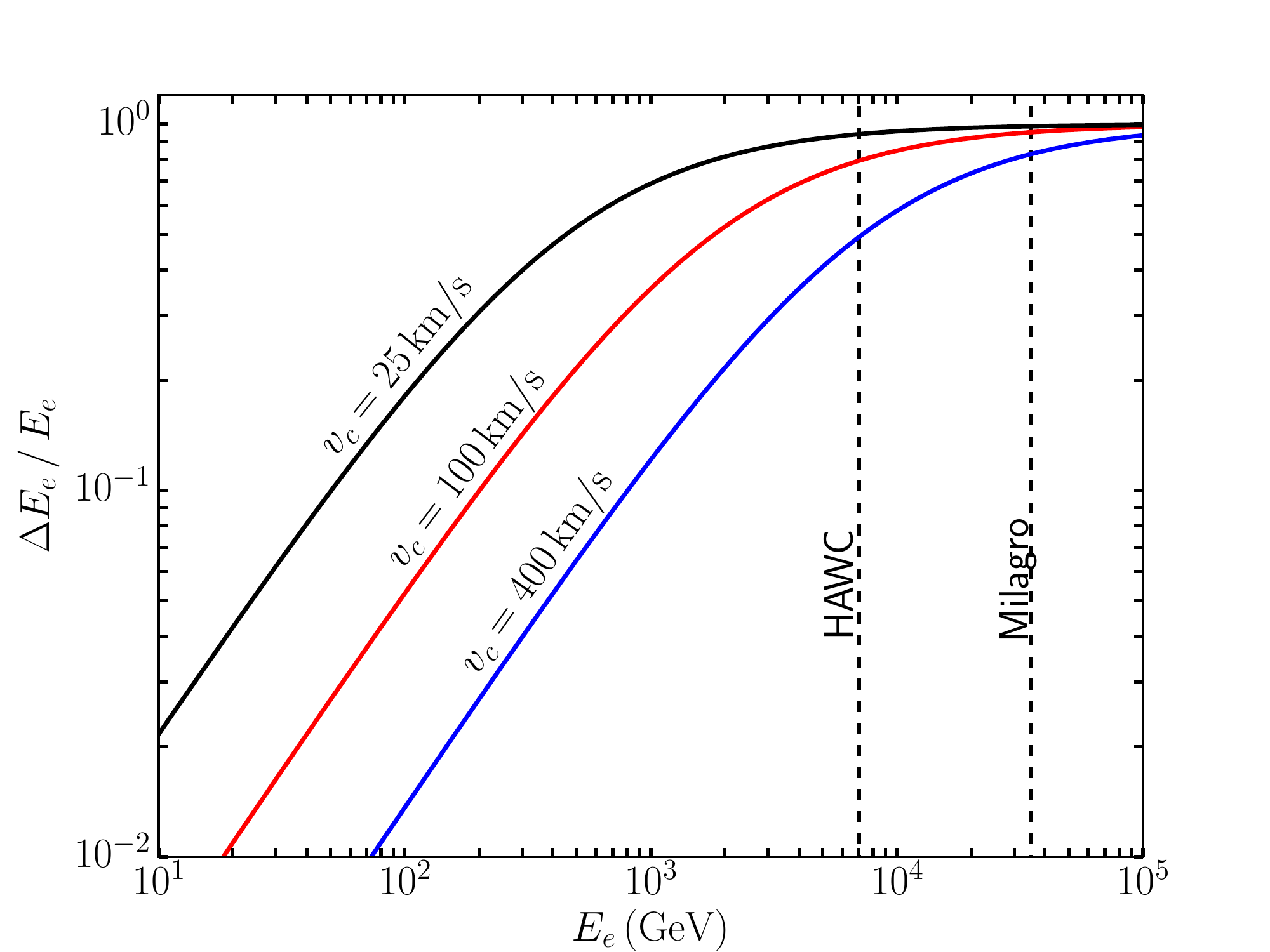}
\hspace{-0.5cm}
\includegraphics[width=3.08in,angle=0]{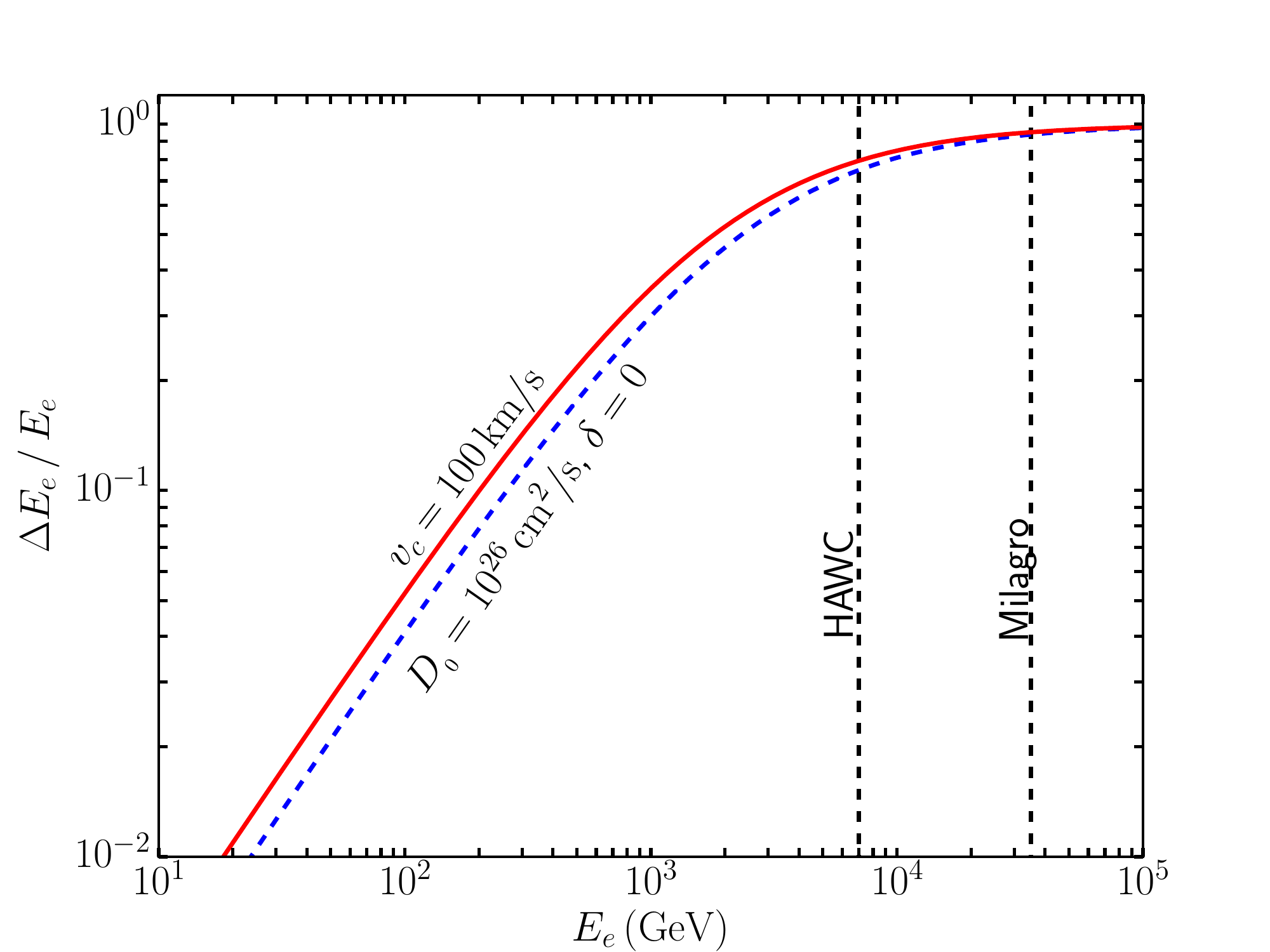}
\caption{Left: The fraction of energy that is lost by an electron over the time that is required to move via convection a distance of 10 parsecs. Right: The mean fraction of energy that is lost by an electron over the time required to move via either convection ($v_c=100$ km/s) or energy-independent diffusion ($D_0=10^{26}$ cm$^2/$s) a distance of 10 parsecs. For comparison, in each frame we show the (gamma-ray) energies at which HAWC and Milagro have reported fluxes.}
\label{convectionloss}
\end{figure}

To address this more quantitatively, we can compare the timescale for energy losses, $\tau_{\rm loss} = 1/E_e b$, to that for the escape from the region, $\tau_{\rm escape} = r_{\rm region}/v_c$. For convective winds with a velocity of $v_c$, particles with $E_e \ll v_c /b r_{\rm region}$ are left largely unaffected, while those with $E_e \gg v_c /b r_{\rm region}$ lose the majority of their energy before escaping. In the left frame of Fig.~\ref{convectionloss}, we plot the fraction of energy that an electron loses before escaping a region of radius 10 parsecs for several values of $v_c$. It is expected that future observations by the Cherenkov Telescope Array (CTA) will provide an important test of this transition by measuring the intensity and angular extent of the $\sim$\,0.1-10 TeV emission from Geminga and B0656+14.

Although we have focused here on a scenario in which convection is responsible for expelling sub-TeV leptons from the regions surrounding Geminga and B0656+14, other means may also be possible. For example, if we simply introduce a diffusion coefficient with no energy dependance ($\delta=0$), we effectively mimic the effects of convection (see the right frame of Fig.~\ref{convectionloss}).

\begin{figure}
\includegraphics[width=3.08in,angle=0]{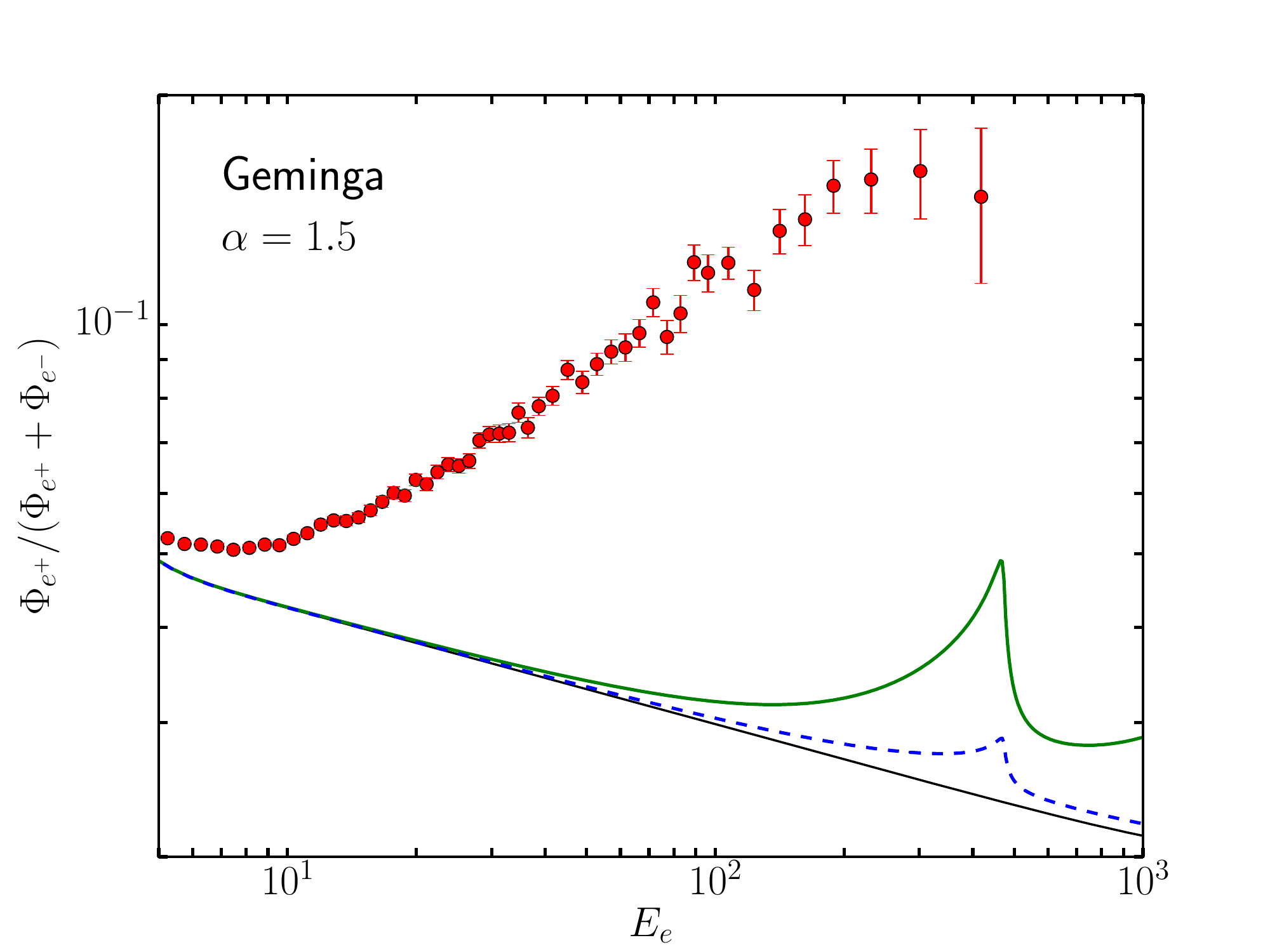}
\hspace{-0.5cm}
\includegraphics[width=3.08in,angle=0]{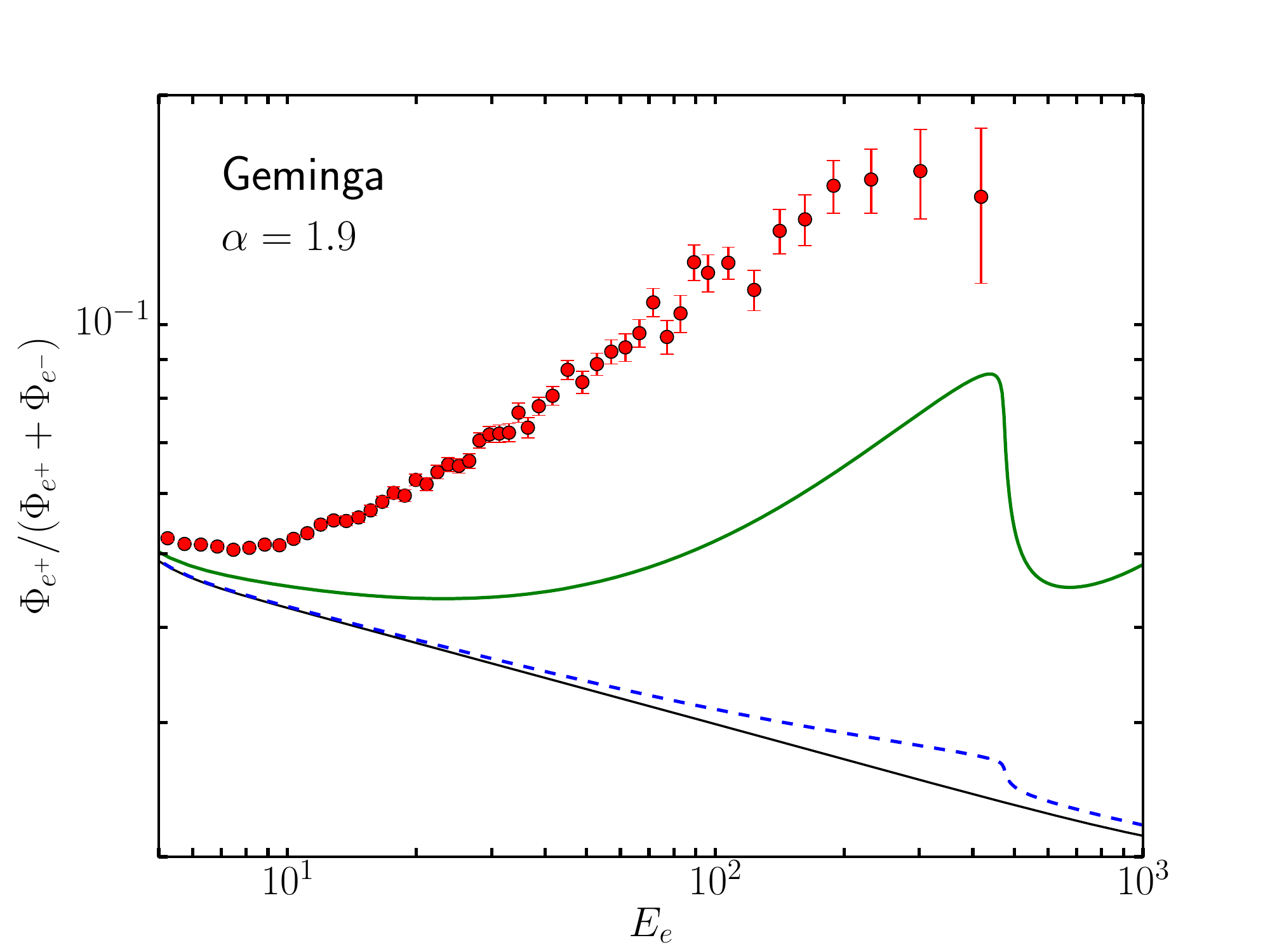}
%
\caption{The cosmic-ray positron spectrum as measured by AMS-02, compared to the predictions from standard secondary production in the ISM (solid black) and including a contribution from the Geminga pulsar. In each case, we have chosen the normalization and spectral shape to match that of the very high-energy gamma-ray emission measured by HAWC and Milagro (see Fig.~\ref{gammageminga}). In the left (right) frame, we adopt a spectrum of injected electrons with an index of $\alpha=1.5$ (1.9), and have in each case selected a value of $E_c$ that provides the best-fit to the combination of the HAWC and Milagro measurements. The solid green curves correspond to a case with strong convection ($v_c =$ 554 km/s $\times \,(r_{\rm region}/10 \, {\rm pc})$). For comparison, we also show as blue-dashed curves the result with weak convection ($v_c =$ 55.4 km/s $\times \,(r_{\rm region}/10 \, {\rm pc})$), although this case is disfavored by the spectral index reported by the HAWC Collaboration.}
\label{posfracgeminga}
\end{figure}

In Fig.~\ref{posfracgeminga}, we plot the cosmic-ray positron fraction as measured by AMS-02, compared to the predictions from the Geminga pulsar, using the same choices of parameters as adopted in Fig.~\ref{gammageminga}. In each frame, the solid black curve denotes the contribution from standard secondary production in the ISM, while the other curves include both this contribution and that from Geminga. We remind the reader that those models with only weak convection (dotted blue curves) do not lead to a spectral index compatible with the measurement of HAWC, and thus should be viewed as a poor fit to the data.

The positron fraction presented in Fig.~\ref{posfracgeminga} includes a distinctive feature at 400-500 GeV, which is a result of energy losses. More specifically, a positron with an infinite energy will be reduced over a time $t$ to an energy of $E_e=(bt)^{-1}$, which for $t=$370,000 years (the age of Geminga) yields a final energy of 476 GeV. Any positrons from Geminga above this energy were injected at later times and thus have not cooled to the same extent. 

The main lesson from the results shown in Fig.~\ref{posfracgeminga} is that when the spectral shape and overall normalization of Geminga are fixed to reproduce the results of HAWC and Milagro, this pulsar is found to generate a non-negligible portion of the observed positron fraction. That being said, the overall size of this contribution to the cosmic-ray positron flux can vary by a factor of order unity depending on the precise values of the convection velocity, $v_c$, and spindown timescale, $\tau$ (see Eq.~\ref{tau}) that are adopted. The impact of the convection velocity is clearly evident in Fig.~\ref{posfracgeminga}. The resulting positron flux scales approximately as $\tau^{-1}$ (we have adopted a value of $\tau=9.1\times 10^3$ years). Furthermore, the time profile of a pulsar's emission could plausibly depart to some extent from that predicted from standard magnetic dipole braking~\cite{Gaensler:2006ua}, potentially altering the normalization of the positron flux predicted here, as well as the inferred age of the pulsar.

\begin{figure}
\includegraphics[width=3.80in,angle=0]{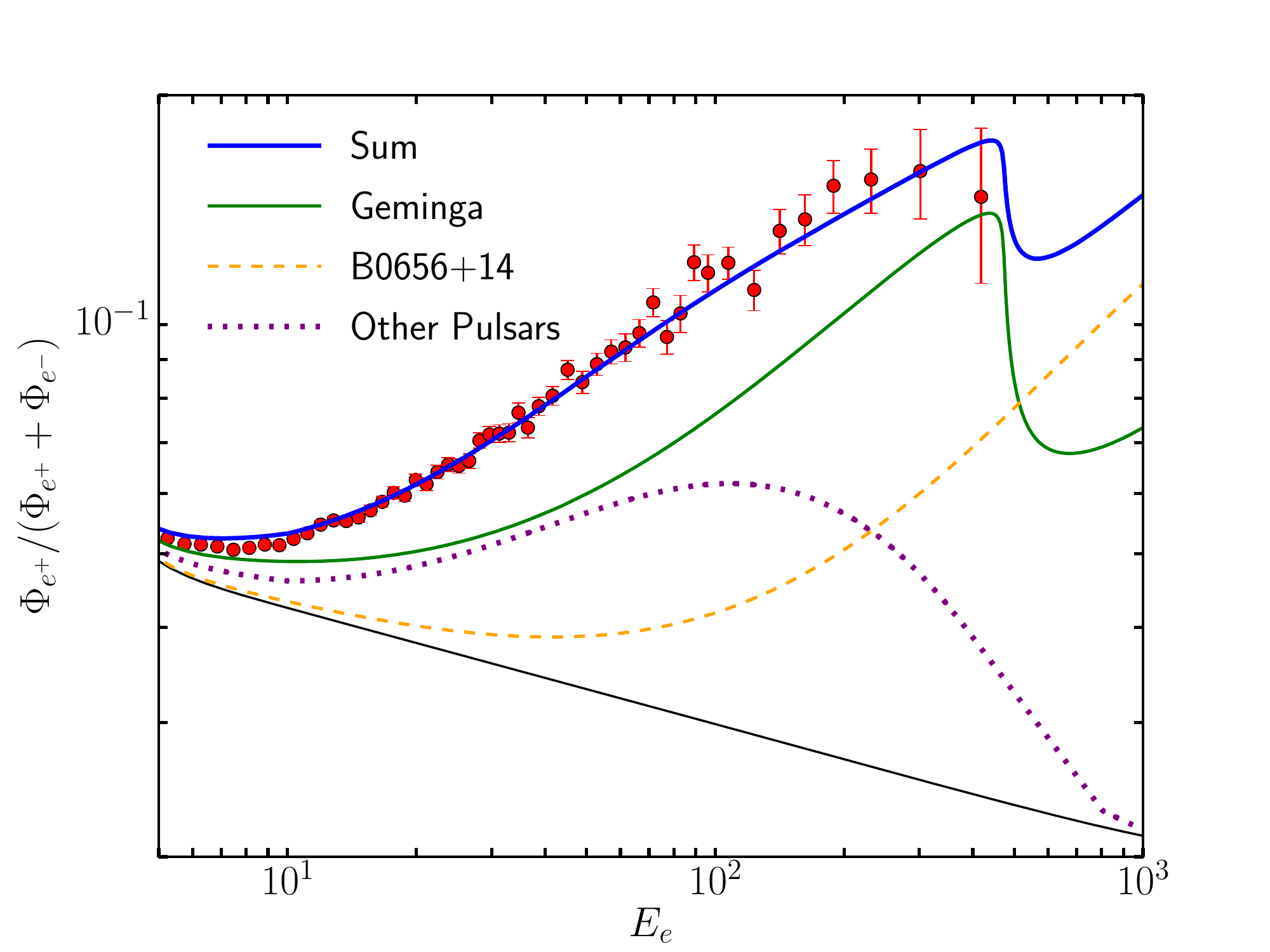}
\caption{As in Fig.~\ref{posfracgeminga}, but showing contributions from Geminga, B0656+14, and from all pulsars more than 0.5 kpc from the Solar System. For each source, we adopted $\alpha=1.9$, $E_{c}=49$ TeV, $v_c=$554 km/s $\times \,(r_{\rm region}/10 \, {\rm pc})$, and normalized their contributions with $\tau=4.3\times 10^3$ years, adopting a total birth rate of two pulsars per century in the Milky Way. While we expect many of these parameters to vary from pulsar-to-pulsar, making a detailed prediction of this kind difficult and possibly unreliable, this calculation provides significant support for the conclusion that a sizable fraction of the observed positron excess originates from pulsars.}
\label{all}
\end{figure}

Finally, in Fig.~\ref{all}, we plot the contributions to the positron fraction from the Geminga and B0656+14 pulsars, as well as the average contribution from those pulsars located more than 500 parsecs away from the Solar System. For each source, we have adopted $\alpha=1.9$, $E_{c}=49$ TeV, $v_c=$554 km/s $\times \,(r_{\rm region}/10 \, {\rm pc})$, and normalized their contributions with $\tau=4.3\times 10^3$ years, and adopting a total birth rate of two pulsars per century in the Milky Way.\footnote{We produce nearly identical results if we instead adopt our default value for $\tau \simeq 9.1\times 10^3$ and a somewhat higher value for the convection velocity, $v_c \simeq $1160 km/s $\times \,(r_{\rm region}/10 \, {\rm pc})$.} For other details regarding the calculation of the contribution from distant pulsars, we direct the reader to Ref.~\cite{Hooper:2008kg}. In reality, we expect many of these parameters to vary from pulsar-to-pulsar, making a detailed prediction of this kind difficult and possibly unreliable. That being said, this exercise clearly demonstrates that in light of the measurements by HAWC and Milagro, it appears very likely that a sizable fraction of the observed positron excess originates from this class of sources. In addition, we note that it was recently shown that the stochastic acceleration of cosmic-ray secondaries in supernova remnants is also likely to contribute to the local positron flux~\cite{Cholis:2017qlb}.

\section{Summary and Conclusions}

In this paper, we have made use of measurements by the very high-energy gamma-ray telescopes HAWC and Milagro to better understand and constrain the injection of high energy electrons and positrons from the nearby pulsars Geminga and B0656+14. The angular extension of the $\gsim$~TeV gamma-ray emission observed from these pulsars indicates that very high-energy leptons are effectively trapped within the surrounding several parsecs around these sources. Furthermore, their very high-energy gamma-ray spectra indicate that lower energy leptons are able to escape more easily, suggesting the presence of strong convective winds, with velocities of several hundred kilometers per second. 

In models that are able to reproduce the characteristics of the gamma-ray emission reported by HAWC and Milagro, these pulsars invariably provide a significant contribution to the local cosmic-ray positron spectrum, and thus to the measured positron fraction. Although it is not yet possible to precisely predict the normalization of the positron flux from these sources, these results show that Geminga and B0656+14 are expected to generate a significant fraction of the observed high-energy positron flux. If we make the entirely reasonable assumption that other pulsars in the Milky Way behave in a similar fashion to Geminga and B0656+14, we find that it is very likely that pulsars are responsible for much, if not the entirety, of the reported positron excess. 

An important test of this conclusion will come from future Imaging Atmospheric Cherenkov Telescopes (IACTs), such as the Cherenkov Telescope Array (CTA). Although existing IACTs have not yet reported any significant detection of TeV-scale emission from Geminga or B0656+14~\cite{Ahnen:2016ujd,2015arXiv150904224A}, next generation telescopes will be far better suited to detect emission that is extended over the angular scales reported by Milagro and HAWC. Such a measurement is expected to be able to confirm the transition to convection-dominated transport at energies below several TeV, and enable us to produce a more detailed determination of the spectrum of electrons and positrons that are injected from this class of sources.
 
\bigskip
\bigskip
\bigskip

\textbf{Acknowledgments.} DH is supported by the US Department of Energy under contract DE-FG02-13ER41958. Fermilab is operated by Fermi Research Alliance, LLC, under contract DE- AC02-07CH11359 with the US Department of Energy. IC acknowledges support from NASA Grant NNX15AB18G and from the Simons Foundation. TL acknowledges support from NSF Grant PHY-1404311.

\bibliography{geminga.bib}
\bibliographystyle{JHEP}

\end{document}